\documentclass[conference]{IEEEtran}
\IEEEoverridecommandlockouts

\usepackage{cite}
\usepackage{amsmath,amssymb,amsfonts}
\usepackage{algorithmic}
\usepackage{graphicx}
\usepackage{textcomp}
\usepackage{xcolor}

\usepackage{seqsplit}

\def\BibTeX{{\rm B\kern-.05em{\sc i\kern-.025em b}\kern-.08em
    T\kern-.1667em\lower.7ex\hbox{E}\kern-.125emX}}
\begin{document}

\title{Chain Reactions: How Nonce Collisions in ECDSA Compromise Polygon MEV Searchers}

\author{\IEEEauthorblockN{
Yash Madhwal\IEEEauthorrefmark{1},
Andrey Seoev\IEEEauthorrefmark{2}, 
Raffaele Della Pietra\IEEEauthorrefmark{2},
Anastasiia Smirnova\IEEEauthorrefmark{2}, 
Yury Yanovich\IEEEauthorrefmark{1}
}
\IEEEauthorblockA{\IEEEauthorrefmark{1}%
  Skolkovo Institute of Science and Technology, Moscow, Russia \\
}
\IEEEauthorblockA{\IEEEauthorrefmark{2}%
  MEV-X, Moscow, Russia 
}
}

\maketitle


\begin{abstract}
ECDSA signatures form the bedrock of blockchain transaction authentication, yet their security critically depends on proper nonce generation. We uncover a critical vulnerability in the Polygon MEV ecosystem: systematic nonce reuse that enables complete private key recovery. Analyzing on-chain data reveals that searchers, driven by the need for sub-second response times in sealed-bid auctions, employ predictable nonce patterns. These patterns create linear relationships between signatures, allowing passive attackers to recover private keys using elementary algebra. We provide a compact linear-system formulation for such attacks, including the dangerous case of cross-wallet nonce collisions, and present concrete evidence of exploitable patterns on Polygon. Our findings demonstrate how protocol-induced latency pressures can lead to catastrophic cryptographic failures in production blockchain systems, where a single implementation error compromises multiple accounts simultaneously.
\end{abstract}


\section{Introduction: The Unforgiving Reality}

Digital signatures are the linchpin of blockchain security, enabling trustless authorization of billion-dollar transactions. ECDSA remains the dominant standard across major cryptocurrency ecosystems, including Polygon's rapidly growing DeFi landscape. In these high-stakes environments, signatures are public, permanent, and high-value targets: a single compromised private key can lead to irreversible financial loss exceeding millions of dollars.

The perennial vulnerability lies in the per-signature nonce~$k$. ECDSA's security model assumes each $k$ is a fresh, unpredictable secret, but practical implementations often falter. Poor randomness, buggy implementations, side-channel leakage, and--most critically for MEV searchers--reused nonces create attack vectors that are actively being exploited. In our analysis of Polygon MEV-related activity, we observe repeated and cross-wallet nonce reuse patterns consistent with unsafe ECDSA signing behavior. These patterns can lead to complete key compromise when exploited. Once exploited, searchers quickly abandon this practice, but the damage is already irreversible.

This work provides both a warning and a mathematical toolkit: we show how nonce compromise reduces to solving small linear systems, enabling attackers to passively harvest private keys from public blockchain data. The threat is not theoretical--it represents a common implementation error with catastrophic consequences.


\section{Background: ECDSA's Achilles' Heel}
Let $(\mathbb{G}, n)$ be an elliptic-curve group of prime order $n$ with generator $G$. A private key is $x \in \{1,\dots,n-1\}$ and the public key is $X = xG$. To sign a message $m$, ECDSA computes a hash $e = H(m)$ (reduced modulo $n$) and samples a per-signature nonce $k \in \{1,\dots,n-1\}$. The signature is the pair $(r,s)$, where
\begin{align}
R &= kG,\qquad r = R_x \bmod n, \nonumber \\
s &\equiv k^{-1}(e + xr) \pmod n.
\label{eq:ecdsa}
\end{align}
Equation~\eqref{eq:ecdsa} reveals the fatal linearity: $s$ depends linearly on private key $x$ once $k$ is fixed. Any structure in $k$ becomes a direct pipeline to $x$.

\section{Threat Model: The Attacker's Playground}
We consider adversaries who monitor public blockchain data--a trivial task in permissionless systems. From observed valid ECDSA signatures $(r_i, s_i)$ on messages $m_i$ with hashes $e_i = H(m_i)$, attackers aim to recover signing keys under three increasingly severe compromise conditions:

\noindent\textbf{C1) Single-wallet nonce reuse.} A signer reuses the same nonce $k$ across multiple signatures. This well-known case has been extensively documented in prior large-scale analyses \cite{Bos2014}.

\noindent\textbf{C2) Linear nonce relations.} Nonces exhibit predictable patterns: $k_i = \alpha_i k_0 + \beta_i$ for known $(\alpha_i,\beta_i)$. Lattice attacks can exploit more complex biases \cite{Breitner2019}.

\noindent\textbf{C3) Cross-wallet nonce collisions (CRITICAL).} Multiple private keys sign with the same nonce $k$, creating interconnected vulnerability chains. Two such collisions can compromise both wallets. This expands the attack surface beyond single-key analysis.

Condition C3 represents an escalating threat: once one wallet is compromised, every collision with it exposes another victim. This creates cascading failures across the ecosystem, making it a particularly dangerous pattern in competitive MEV environments.

\section{Linear-System Formulation: The Mathematics of Key Theft}
From~\eqref{eq:ecdsa}, each signature yields:
\begin{equation}
x r_i - s_i k_i \equiv -e_i \pmod n.
\label{eq:lin2}
\end{equation}
This linear relationship becomes a weapon when $k_i$ exhibits structure.

\subsection{Case C1: The Classic Catastrophe}
If two signatures share nonce $k$, their $r$ values match (with overwhelming probability). Let $(r, s_1)$ sign $e_1$ and $(r, s_2)$ sign $e_2$ with the same $k$. Subtracting equations eliminates $x$:
$(s_1 - s_2)k \equiv (e_1 - e_2) \pmod n$,
yielding:
$k \equiv (e_1 - e_2)(s_1 - s_2)^{-1} \pmod n$, and
$x \equiv (s_1 k - e_1) r^{-1} \pmod n$.
One reuse, complete compromise.

\subsection{Case C2: The Pattern Problem}
When nonces satisfy $k_i = \alpha_i k_0 + \beta_i$ with known coefficients, substituting into~\eqref{eq:lin2} gives:
\begin{equation*}
s_i(\alpha_i k_0 + \beta_i) \equiv e_i + x r_i \pmod n.
\end{equation*}
Rearranging yields linear equations in $(x, k_0)$:
\begin{equation*}
r_i x - (s_i \alpha_i) k_0 \equiv s_i\beta_i - e_i \pmod n.
\end{equation*}
Collect $t$ signatures to build the solvable system:
\begin{equation*}
A \cdot
\begin{bmatrix}
x\\
k_0
\end{bmatrix}
\equiv b \pmod n.
\end{equation*}

\subsection{Case C3: Cross-Wallet Chain Reactions (Polygon's Reality)}
Consider two private keys $d_A$ and $d_B$ that both sign with nonce $k_1$ (producing $r_1$). This gives equations:
\begin{align*}
s_{A1} &\equiv k_1^{-1}(e_{A1} + r_1 d_A) \pmod n, \\
s_{B1} &\equiv k_1^{-1}(e_{B1} + r_1 d_B) \pmod n.
\end{align*}
Eliminating $k_1$ yields one linear relation between $d_A$ and $d_B$. Add a second collision with nonce $k_2$ (producing $r_2$), we now have two independent linear equations--exactly enough to solve for both unknowns. The matrix formulation:
\begin{equation}
\begin{bmatrix}
s_{B1} r_1 & -s_{A1} r_1 \\
s_{B2} r_2 & -s_{A2} r_2
\end{bmatrix}
\begin{bmatrix}
d_A \\ d_B
\end{bmatrix}
\equiv
\begin{bmatrix}
s_{A1} e_{B1} - s_{B1} e_{A1} \\
s_{A2} e_{B2} - s_{B2} e_{A2}
\end{bmatrix}
\pmod n.
\end{equation}
\textbf{Critical implication:} Two cross-wallet collisions compromise both wallets completely. This pattern, which can be modeled as a bipartite graph of keys and nonces \cite{Brengel2018}, creates an ecosystem-wide risk chain.

\section{Case Study: Nonce Reuse Patterns in Polygon MEV Activity}
The shift from spam-based Priority Gas Auctions to sealed-bid FastLane auctions (Polygon Atlas) created extreme time pressure: bids must be submitted within~250 ms windows. This drives searchers to reuse nonce values ($k$) as a latency optimization, sacrificing cryptographic security for microseconds.

Using custom Python tools to scan a representative sample of Polygon blocks, we observe three failure modes in transaction logs: constant $k$ values, sequential reuse, and cross-wallet collisions. While abandoned post-exploitation, historical trails persist on-chain and reveal a pattern of high-speed, low-entropy nonce generation. This aligns with historical incidents where poor random number generation led to key compromise \cite{AndroidBug2013, BlockchainBug2013}.

\noindent\textbf{Collision 1 (same k):}\\
\seqsplit{0xc1e972bbe128937a76edb3a5595d3fdf3e6fbcd7029fce1916607ce0bb9073fd}\\
\seqsplit{0x765aab214b5e61ae75d957c2b7200221a6ea2664e1386ebd7948920ab9662200}

\noindent\textbf{Collision 2 (same k):}\\
\seqsplit{0x722f948b95149cb060f951a8daad748936e9a2664fc3e455b8fcb4e74d9d2666}\\
\seqsplit{0xd2ad403e776f38001bb59dbd7b04580789e3b5d456d9511136c922aabb35448f}

These create chain reaction vulnerabilities: compromising one searcher's key exposes every collided wallet. Our Python implementation confirms complete key recovery. This case study illustrates how latency-sensitive blockchain workflows can coexist with unsafe nonce generation practices, leading to exploitable vulnerabilities.

\section{Mitigations: Preventing Common Errors}
Several mitigations can address these common errors. Immediate actions include switching to RFC 6979 \cite{rfc6979} for deterministic nonce generation, which eliminates randomness failures; continuous monitoring for repeated $r$ values across wallets; immediate key rotation after any signing environment change; and historical audits of past transactions for vulnerability patterns.

Long-term solutions focus on wallet hardening through the auditing of signing components for entropy quality, developer education emphasizing nonce security in blockchain development curricula, protocol upgrades to consider transitions to Schnorr or EdDSA with better nonce handling, and industry-wide enforcement of best practices for secure signing implementations.

\section{Conclusion: Learning from Common Mistakes}
This paper demonstrates how ECDSA nonce compromise--particularly cross-wallet collisions--enables trivial private key recovery through linear algebra. These attacks highlight a common implementation error that persists despite being well-understood in cryptographic literature. Our linear-system formulation provides both a warning and a diagnostic tool, emphasizing that developers must treat nonce generation with the seriousness it demands. To facilitate further research and tooling, we will make our analysis and key recovery scripts publicly available upon acceptance.

The cryptography is unforgiving: a single implementation error can lead to total financial loss. As blockchain systems continue to handle increasing value, we must move beyond recognizing these vulnerabilities to systematically preventing them through education, better tooling, and industry-wide standards. The alternative is a continuous cycle of exploitation and loss in an ecosystem where speed often trumps security.

\bibliographystyle{IEEEtran}

\end{document}